\documentclass[12pt,equations,fleqn]{article}
\usepackage{epsf}
\begin{document}
\begin{titlepage}
\def\baselinestretch{1.4}
\null
\begin{center}
{\LARGE
Phase Diagram of a Loop on the Square Lattice }
\vskip 10mm
{\Large Wenan Guo~$^{\S}$\footnote{present address: Physics Department,
Beijing Normal University, Beijing 100875, P. R. China},
Henk W.J. Bl\"{o}te~$^{\S \dag}$\footnote{e-mail: bloete@tn.tudelft.nl} \\ 
and Bernard  Nienhuis~$^{\ddag}$ \\}
\vskip 5mm
{\em
$^{\S}$ Faculty of Applied Science,  P.O. Box 5046,
2600 GA Delft, The Netherlands \\
$^{\dag}$ Instituut Lorentz,
Universiteit Leiden, Niels Bohrweg 2,
Postbus 9506, 2300 RA Leiden, The Netherlands\\
$^{\ddag}$ Instituut voor Theoretische Fysica,
Universiteit van Amsterdam, Valckenierstraat 65,
1018 XE Amsterdam, The Netherlands\\}
\end{center}
\vskip 2mm
\def\baselinestretch{1.6}
\begin{abstract}
The phase diagram of the O($n$) model, in particular the special case
$n=0$, is studied by means of transfer-matrix
calculations on the loop representation of the O($n$) model. The model
is defined on the square lattice; the loops are allowed to collide at
the lattice vertices, but not to intersect. The loop model contains
three variable parameters that determine the loop density or temperature,
the energy of a bend in a loop, and the interaction energy of colliding
loop segments. A finite-size analysis of the transfer-matrix results 
yields the phase diagram in a special plane of the parameter space.
These results confirm the existence of a multicritical point and an
Ising-like critical line in the low-temperature O($n$) phase.
\end{abstract}
\vfill
{\em Keywords:} O($n$) model; Polymers; Phase diagram.
\end{titlepage}

\section{Introduction }
The O($n$) model is originally defined as a model of $n$-component spins,
with O($n$) symmetric interactions. This model appears to be meaningful
only when $n$ is a positive integer. However, a mapping \cite{dG}
has been found on a model of closed loops in which $n$ appears as the 
Boltzmann weight associated with each loop: hence $n$ has become a
continuously variable parameter. The critical point, which is
characterized by  e.g. the divergence of the susceptibility in the
spin representation, corresponds with a diverging loop size in the
loop model. The latter model, in particular in two dimensions, is
accessible by exact and numerical analyses, and is thus also relevant
for the exploration of the O($n$) spin model. Indeed, a wealth of
results have been found for the O($n$) loop model. For
instance, an exact critical line was found for an O($n$) model on the
honeycomb lattice, as well as the associated thermal and magnetic
exponents \cite{3Nhon,Baxh,3BB,JSuzu}. The
critical line covers the interval $-2\leq n \leq 2$ of the loop model
and thus applies to $n=1$ and $n=2$ in the spin model. Another special
case of the O($n$) loop model is $n=0$, where the weight of a loop
vanishes. This does not mean that the system is empty: the O(0) model 
may be formulated in terms of the derivative of the O($n$) partition
function to $n$, taken at $n=0$ so that one loop remains, finite or
infinite. Thus the nonintersecting O($0$) loop model can be applied
to describe self-avoiding walks (SAW's) or polymers. Furthermore
the $n=-2$ case corresponds to random walks without excluded-volume
interactions \cite{BT,MEF}.

More complicated phase behavior is observed when attractive interactions
between the loop segments are introduced. As long as these attractions
are sufficiently small, the O($0$) ordering transition remains in the
same universality class. But when the interactions reach a certain
threshold a tricritical point appears, called the theta point
\cite{DS}, and at even stronger attractions the phase transition turns
first order. While exact information is scarce in the temperature
versus attraction phase diagram, a numerical analysis \cite{bbn}
confirms the picture sketched here.

This tricritical theta point was first found \cite{DS} for a loop model 
with vacancies on the honeycomb lattice. The same type of transition,
describing a collapsing polymer, was found for the square lattice O($0$)
loop model, namely on branch 0 of Ref.~\cite{bn}.
The latter model did, in addition to a
temperature variable and an attraction between loop segments, also 
contain a parameter that distinguishes between straight loop segments
and 90 degree bends. The results obtained for the square lattice included
a generalization \cite{bn} of the theta transition to values $n \neq 0$.
Remarkably, the enlarged parameter space of three parameters (not 
counting $n$) appears to contain also  a different type of collapse
transition \cite{bn,3WBN} for $n=0$. Like the theta point found on the
square lattice, it generalizes to other values of $n$ (branch 3 of
Ref.~\cite{bn}). 

To understand this second collapse mechanism, it may help to identify
straight loop segments in terms of Ising-like degrees of freedom in
the O($n$) model \cite{bn,bbn}. When
the straight loop segments are suppressed, Ising-like order appears.
Thus the newly observed multicritical point has been identified
as the point along the O($n$) critical line where the Ising degrees of
freedom become critical. These new effects are a consequence of the 
underlying lattice structure and are thus not generally expected in a
model of free polymers. However, they could be present in the case of a
polymer influenced by a periodic potential caused by a crystalline
substrate.

The freezing of these Ising degrees of freedom does not only occur
on the line of phase transitions, but is also expected in the
low-temperature O($n$) phase. Thus an Ising-like transition would
extend from the multicritical point into the
dense phase of the loop model.  Indeed, the results reported in
Ref.~\cite{bn} include a branch of critical points where the critical
exponents reflect both the Ising and the low-temperature O($n$) universal
critical behavior.

These observations have been interpreted in terms of a conjectured and
qualitative phase diagram \cite{bn}. In the present paper we confirm the 
validity of this phase diagram, including the predicted Ising line, by
means of finite-size scaling and a transfer-matrix technique.
Our investigations concern the $n=0$ case of the O($n$) model on
the square lattice as defined in Ref.~\cite{bn}. We briefly recall the
definition of the partition function
\begin{equation}
    Z_{\rm loop}=\sum_{\cal L}n^{N_l}x^{N_x}y^{N_y}z^{N_z}
\end{equation}
which depends on three vertex weights $x$, $y$, and $z$, associated with
a 90-degree bend of a loop segment, a straight loop segment, and a vertex
visited by two 90-degree bends: a collision. Intersections are forbidden
and a weight 1 is assigned to an empty vertex (not visited by a loop). The
exponents $N_x$, $N_y$ and $N_z$ denote the total number of vertices of
the type indicated. The number of closed loops is  $N_l$, and thus, after
differentiation with respect to $n$, only loop configurations ${\cal L}$
with $N_l=1$ contribute to the partition function of the $n=0$ model.
When the weight $y$ of the straight loop segments is set to 0, we freeze
the Ising-like degrees of freedom mentioned above. Then the model reduces
to that investigated in Ref.~\cite{bbn} and displays a theta point. The
special character of the $y=0$ case also becomes apparent from the fact
that the model, at least for even system sizes, is symmetric with respect
to a change of sign of $y$. The point $y=0$ is special on the basis of
this symmetry. In this paper we explore nonzero values of $y$ and 
restrict ourselves to the $x=y$ plane. The weight $x=y$ plays the role of
a temperature-like parameter (at least when $z$ is small) and drives the
SAW-type transition between the vacuum and the dense phase.
The role of the parameter $z$ is
to adjust the attractive interaction between loop segments.

The rest of this paper is composed as follows.
In Section 2 we summarize the numerical procedures and in Section 3
we present the results and discuss the phase diagram thus obtained.

\section{Numerical procedures }

We consider the square O(0) model, wrapped on an infinitely long cylinder
of circumference $L$, the finite size of the system, which is measured 
in elementary lattice spacings. One can derive quantities of interest,
such as the free energy and correlation lengths, from the eigenvalues of
the transfer matrix. For this purpose we make use of the construction
of the O($n$) loop model transfer matrix given in Ref.~\cite{bn}.

The relevant eigenvalues of the transfer matrix are obtained by a method
explained e.g. in Ref.~\cite{bnigh}. This method is based on iterative 
multiplication of a vector by the transfer matrix. Even when $n=0$ this
procedure generates non-empty configurations containing a part of a loop.
Closure of such a loop leads to a zero contribution to the resulting
vector, but there will always be contributions due to configurations
in which the loop is not (yet) closed. These configurations contribute
to the eigenvalues of the transfer matrix. Therefore the transfer matrix
allows the exploration of non-empty configurations of the O(0) model.

The O($n$) spin-spin correlation function can simply be translated into
the language of the O($n$) loop model, as the ratio $Z'/Z$, where $Z$
is the partition sum of the loop  model and $Z'$ a similar sum, but 
restricted to loop configurations that include a non-closed loop
segment connecting
the two correlated spins. For spins separated along the length direction
of the cylinder, we may obtain $Z'$ from the `odd transfer matrix' that 
describes configurations which include such an additional single segment
\cite{bn}. The partition zum $Z$ is obtained from the `even transfer
matrix' that allows only configurations in which loop segments are 
pairwise connected at the end of the cylinder. This is a convenient
method to derive the magnetic correlation length and its associated
scaling dimension. 
For further details see Ref.~\cite{bn}.

The relation between the magnetic correlation length $\xi_h$, the
largest eigenvalue $\Lambda^{(0)}_L$ of the even transfer matrix and
the largest eigenvalue $\Lambda^{(1)}_L$ of the odd transfer matrix is
\begin{equation}
\xi_h(L)= 1 / \ln(\Lambda^{(0)}_L/\Lambda^{(1)}_L)
\end{equation}
The finite-size scaling behavior of $\xi_h$ and the principle of 
the `phenomenological renormalization' \cite{MPN} of $\xi_h$ provide a
method to explore the phase diagram of the model. For this purpose we
define the `scaled gap' as $X_h(t,L)=L/2\pi\xi_h(t,L)$. 
Apart from the finite size $L$, it depends on a temperature-like 
scaling field $t$. More than one scaling field can play a role but we
consider only one field at a time.

For conformally invariant models, the scaled gaps $X_h(0,L)$ tend to the
magnetic scaling dimension $X_h$ for $L \rightarrow \infty$ \cite{Cardy-xi}.
For nonzero but small $t$, we expect finite-size scaling 
behavior according to
\begin{equation}
X_{h}(t, L)=X_h+1/(2\pi) tL^{y_t}[d\xi_h^{-1}(t,1)/dt]_{t=0}+\cdots
\label{scx}
\end{equation}
where $y_t$ the renormalization exponent associated with $t$. 
If the temperature-like field is irrelevant ($y_t <0)$, 
$X_h(t,L)$ converges to $X_h$ with increasing $L$. If it is relevant,
$X_h(t,L)$ diverges but for small $t$ Eq.~(\ref{scx}) may still give
a good description of the finite-size data within the range of 
available $L$ values. This formula serves
well for the analysis of the finite-size results $X_h(t,L)$
in terms of the phase diagram.

\section{Transfer matrix analysis}

We restrict our parameter space by setting $x=y$. A preliminary scan of
the phase diagram in the resulting plane, based on the leading eigenvalues
and eigenvectors for small $L$, confirmed the existence of three distinct
phases:
\begin{enumerate}
\item
A high-temperature or vacuum phase for small $x$ and $z$. It corresponds
with the paramagnetic phase in the O($n$) spin language. The vacuum
state is separated by a gap from non-empty states.
\item
A moderately dense phase for large $x$ but not too large $z$. Type-$y$
vertices are abundant, suggesting Ising disorder as described in
Ref.~\cite{bn}. This is the gapless low-temperature phase of the O($n$)
spin model.
\item
At sufficiently large $z$ we observe an even denser phase, dominated by
type-$z$ vertices. Type-$y$ vertices are scarce, suggesting Ising order.
This phase is of the low-temperature O($n$) type, but there exists a gap
between the leading eigenvalues of the even and odd transfer matrices
that does not vanish when $L \to \infty$.
This is because, on lattices with even $L$, loop configurations composed
of only type-$z$ vertices, occur only in the case of the even transfer
matrix. The odd transfer matrix is unable to cover an even $L$ lattice 
completely with type-$z$ vertices.

\end{enumerate}
In order to locate lines of phase transitions separating these 
three regions, we set $z$ at some fixed value, and adjust $x$ to its
transition value $x_c$. We rewrite Eq.~(\ref{scx}) as a function of $x$,
where we may e.g. take  $t=x-x_c$ in first order to represent the relevant
temperature field that vanishes at the transition. This leads to
\begin{equation}
X_h(x,L)=X_h(x,L+2)
\label{sceq}
\end{equation}
where we neglect corrections due to irrelevant fields. Thus numerical
solution of $x$ of Eq.~(\ref{sceq}) yields an estimate of $x_c$.
Solutions were obtained for even system sizes $L=2 \cdots 10$ in
Eq.~(\ref{sceq}). We ignored the odd systems 
because they display additional effects due to odd-even alternation.

\begin{figure}
\epsfxsize=120mm
\epsffile{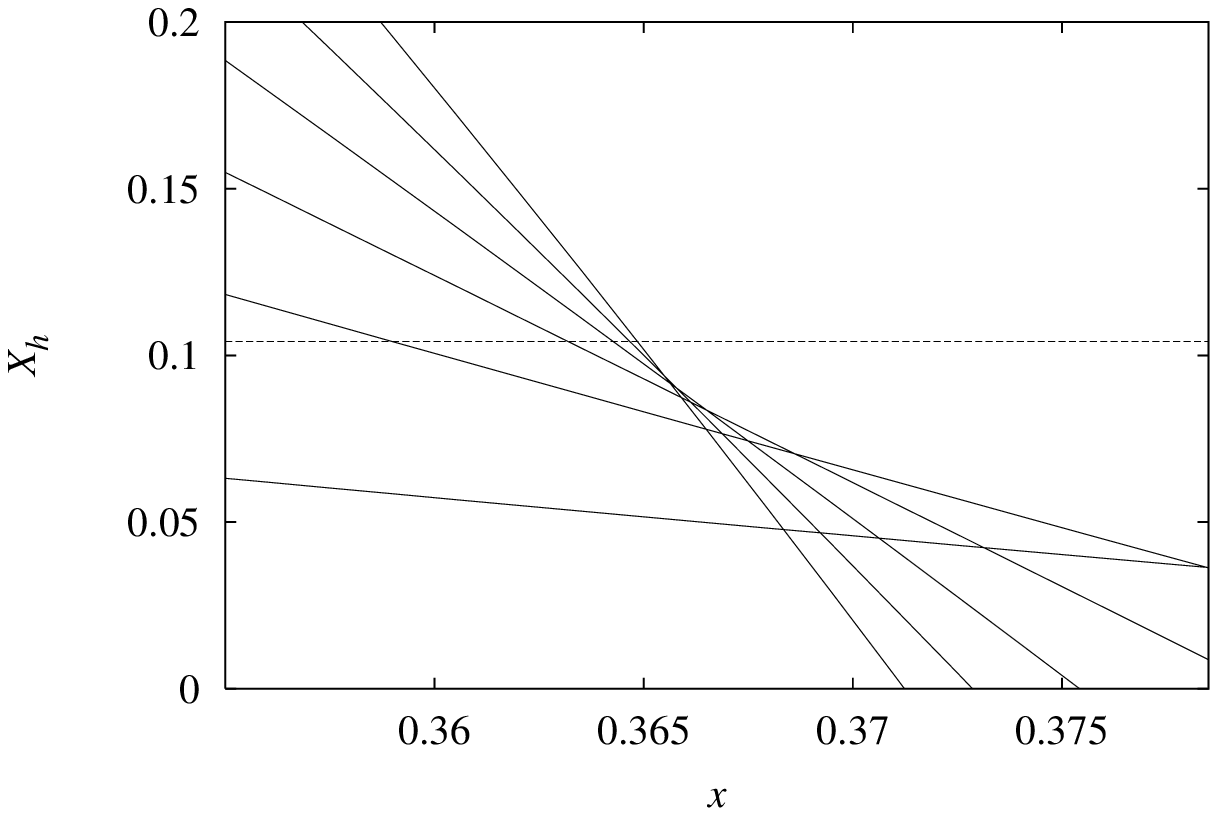}
\vskip 5mm
\caption{ Scaled magnetic gap as a function of the bond weight $x$ 
at constant $z=0.3$. Different curves apply to different
finite sizes $L=2$, 4, 6, 8, 10, and 12. The steeper 
the line, the larger $L$. The scaled gaps at the intersections are
seen to approach the expected value $X_h=5/48$ (dotted line).
}
\label{xpl1}
\end{figure}

\begin{figure}
\epsfxsize=120mm
\epsffile{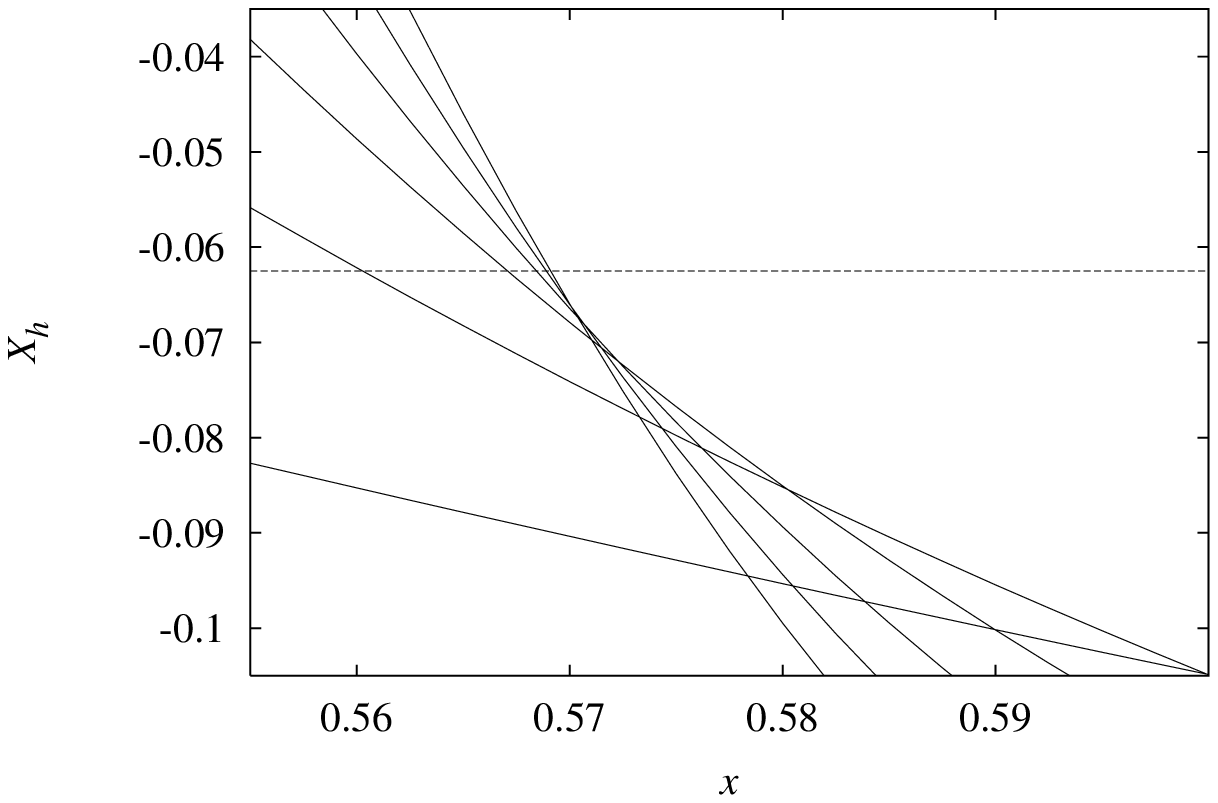}
\vskip 5mm
\caption{ Scaled magnetic gap as a function of the bond weight $x$  
at constant $z=0.8$. Different curves apply to different finite
sizes $L=2$, 4, 6, 8, 10, and 12. The steeper 
the line, the larger $L$. The scaled gaps at the intersections are
seen to approach the expected value $X_h=-1/16$ (dotted line).
}
\label{xpl2}
\end{figure}

\begin{figure}
\epsfxsize=120mm
\epsffile{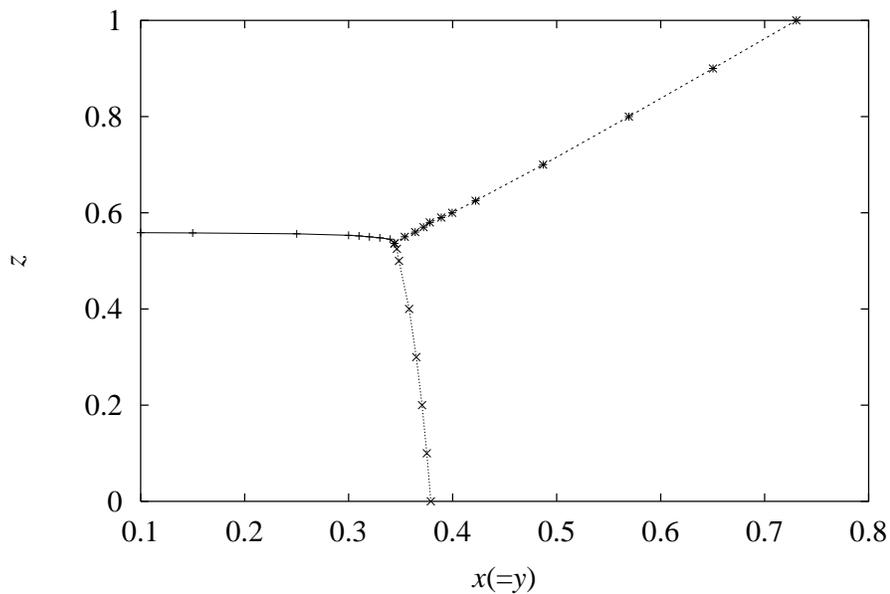}
{\Large
\caption{ Phase diagram of the O($0$) model for $x=y$. The vertical scale 
parametrizes the strength of the attractions between loop segments. The
horizontal scale indicates the weight of vertices visited once by a loop.
The data points were numerically obtained; their errors are much smaller
than the symbol sizes. The curves serve only as a guide to the eye.}
}
\label{phdia}
\end{figure}

Figures  \ref{xpl1} and  \ref{xpl2} illustrate how these $x_c$ 
estimates are obtained. They show the scaled gaps for various system
sizes as a function of $x$, for $z=0.3$ (Fig.  \ref{xpl1}) and for
$z=0.8$ (Fig.  \ref{xpl2}). Intersections of the curves for successive
values of $L$ yield estimates of $x_c$. The intersections do not coincide
as a consequence of corrections to scaling, but seem to converge well
with increasing size $L$.
The expected corrections behave as negative powers of $L$ and
extrapolation of the $x_c$ estimates was done accordingly. The results
are shown in Fig. \ref{phdia}.

Another way to investigate the phase diagram is to fix $x=y$ at some
value and to vary $z$. For small values of $x$ the leading eigenvalues
of the even transfer matrix display intersections, leading to a kink
in the free energy when $z$ is increased: the first-order phase
transition between the vacuum phase and the dense phase, as already 
expected in Ref.~\cite{bn}.
The intersections provide a convenient way to solve for the location
of the first-order line: one just solves for the value of $z$ where the
two largest eigenvalues become degenerate. This procedure leads to
results that seem to converge well; they are included in Fig. \ref{phdia}.
The data shown in this figure are in a full (but, of course, only
qualitative) agreement with the conjectured
phase diagram of Ref.~\cite{bn}.

The solutions of Eq.~(\ref{sceq}) do not only yield estimates of the 
critical point, but also for the corresponding magnetic scaling dimension
$X_h$ which relates to the exponent $\eta$ describing the decay of the 
magnetic correlation function as $\eta=2 X_h$.
This allows us to verify the expected universal behavior along the
critical lines. Indeed, as is already illustrated in Fig. \ref{xpl1}, 
the finite-size estimates of $X_h$ for the
critical line at $z < 0.4$ are in a good agreement with the exactly 
known value $X_h = 5/48 = 0.104166\cdots$ for the ordinary O(0) or SAW
critical point.
Furthermore, for $z > 0.6$ the $X_h$ estimates seem to converge 
well to the expected value \cite{bn} $X_{h} = -1/16 $ for branch 4, which
is the sum of the low-temperature magnetic exponent (-3/16) and the Ising
exponent (1/8).
The crossover between these different ranges of $z$ appears to take
place near $z = 0.54$. At this value, the $X_h$ estimates behave in
a way consistent with $X_h=-5/48$ as found for the $n=0$ point of
branch 3 \cite{3WBN}.

These results confirm that the unusual universal behavior reported for
branches 3 and 4 \cite{bn} is not restricted to the special points 
that admit exact solutions, but apply to larger parameter spaces,
at least for $n=0$. In the $x=y$ plane we locate
one point corresponding with branch 3, and a line of points that
display branch-4 like behavior. Since the $n=0$ points of
branches 3 and 4 do not lie in the $x=y$ plane, it appears that 
the universality class of branch 3 covers a 1-dimensional manifold
of the $x,y,z$ parameter space, and that the  universality class of
branch 4 covers a 2-dimensional manifold. It seems reasonable to expect
that a similar picture exists for $n\neq 0$, at least for a range of
$n$ values. This will be the subject of further investigations.

The calculations reported here are modest in size: they required only
a few hours of CPU time on a Silicon Graphics workstation and a few
hours on a personal computer.

{\em Acknowledgements}: 
This research is supported in part by the FOM
('Stichting voor Fundamenteel Onderzoek der Materie') which is
financially supported by the NWO ('Nederlandse Organisatie voor
Wetenschappelijk Onderzoek').

\newpage

\end{document}